\documentstyle[12pt,a4]{article}

\input{amssym.def}
\input{amssym.tex}

\newfont{\Bbbtw}{msbm10 scaled 1200}
\newfont{\fraktw}{eufm10 scaled 1200}
\newfont{\ninBbb}{msbm9} \newcommand{\sBbb}[1]{\mbox{\ninBbb #1}}
\newfont{\eigBbb}{msbm8} 
\newfont{\sevBbb}{msbm7} 
\newfont{\ninfrak}{eufm9} \newcommand{\sfrak}[1]{\mbox{\ninfrak #1}}
\newfont{\eigfrak}{eufm8} \newcommand{\ssfrak}[1]{\mbox{\eigfrak #1}}
\newfont{\sevfrak}{eufm7} 

\newcommand{\smin}{\,\raisebox{0.06em}{${\scriptstyle \in}$}\,}
\newcommand{\LSA}{Lie superalgebra}

\newcommand{\ca}{\mbox{\fraktw a}}
\newcommand{\cg}{\mbox{\fraktw g}}
\newcommand{\ch}{\mbox{\fraktw h}}
\newcommand{\ck}{\mbox{\fraktw k}}
\newcommand{\cn}{\mbox{\fraktw n}}
\newcommand{\cz}{\mbox{\fraktw z}}

\newcommand{\Wedge}{\raisebox{0.5ex}{${\scriptstyle \bigwedge}$}}
\begin{document}

\makebox[13.5cm][r]{}%{preliminary draft}

\vspace{1.5cm}

\begin{center}
{\bf Lie Superalgebras and the Multiplet Structure of the Genetic Code I: \\
     Codon Representations}

\vspace{1.5cm}

 {\large Michael Forger and Sebastian Sachse
         \footnote{Work supported by FAPESP (Funda\c c\~ao de Amparo
                   \`a Pesquisa do Estado de S\~ao Paulo) and CNPq
                   (Conselho Nacional de Desenvolvimento Cient\'{\i}fico
                   e Tecnol\'ogico), Brazil}}
 \\[5mm]
 {\em Departamento de Matem\'atica Aplicada, \\
      Instituto de Matem\'atica e Estat\'{\i}stica, \\
      Universidade de S\~ao Paulo, \\
      Cx.\ Postal 66281, BR--05315-970 S\~ao Paulo, S.P., Brazil} \\[1cm]
 {\large \today}
\end{center}

\thispagestyle{empty}

\vspace{1cm}

\begin{abstract}
\noindent
It has been proposed \cite{hor} that the degeneracy of the genetic code,
i.e., the phenomenon that different codons (base triplets) of DNA are
transcribed into the same amino acid, may be interpreted as the result
of a symmetry breaking process. In ref.\ \cite{hor} this picture was
developed in the framework of simple Lie algebras. Here, we explore the
possibility of explaining the degeneracy of the genetic code using basic
classical Lie superalgebras, \mbox{whose} representation theory is sufficiently
well understood, at least as far as typical representations are concerned.
In the present paper, we give the complete list of all typical codon
representations (typical $64$-dimensional irreducible representations),
whereas in the second part, we shall present the corresponding branching
rules and discuss which of them reproduce the multiplet structure of the
genetic code.
\end{abstract}

\vfill

\pagebreak       

\section{Introduction}

The discovery of the molecular structure of DNA by Watson and Crick in 1953
was the most important step towards an understanding of the physiological
basis for the storage and transfer of genetic information. DNA is a
macromolecule in the form of a double helix which encodes this information
in a language with 64 three-letter words built from an alphabet with a set
of four different letters (the four nucleic bases attached to the backbone
of a DNA molecule). These words are called codons and form sentences called
genes. Each codon can be translated into one of twenty amino acids or a
termination signal. This leads to a degeneracy of the code in the sense
that different codons represent the same amino acid, that is, different
words have the same meaning. In fact, the codons which code for the same
amino acids form multiplets as follows:
\begin{tabbing} 
 $\bullet$ \quad 5 quadruplets \qquad \= \kill
 $\bullet$ \quad 3 sextets            \> Arg, Leu, Ser \\[1ex]
 $\bullet$ \quad 5 quadruplets        \> Ala, Gly, Pro, Thr, Val \\[1ex]
 $\bullet$ \quad 2 triplets           \> Ile, Term \\[1ex]
 $\bullet$ \quad 9 doublets           \> Asn, Asp, Cys, Gln, Glu,
                                         His, Lys, Phe, Tyr \\[1ex]
 $\bullet$ \quad 2 singlets           \> Met, Trp
\end{tabbing}
When a protein is synthesized, an appropriate segment of one of the two
strings in the DNA molecule (or more precisely, the mRNA molecule built
from it) is read and the corresponding amino acids are assembled
sequentially. The linear chain thus obtained will then fold to the
final configuration of the protein.

These well-known facts, however, provide no explanation as to why just
this special language has been chosen by nature. Since its discovery,
the genetic code has essentially remained a table connecting codons
(base triplets) with the amino acids they represent, but a complete
understanding of its structure is still missing.

A new approach to the question was suggested in 1993 by Hornos \& Hornos
\cite{hor} who proposed to explain the degeneracy of the genetic code as
the result of a symmetry breaking process. The demand of this approach can
be compared to explaining the arrangement of the chemical elements in the
periodic table as the result of an underlying dynamical symmetry which is
reflected in the electronic shell structure of atoms. Another comparable
example is the explanation of the multiplet structure of hadrons as a result
of a ``flavor'' $SU(3)$ symmetry, which led to the quark model and to the
prediction of new particles. An interesting and important feature of this
``flavor'' symmetry is its internal or dynamical nature, that is, it is an
internal property of the dynamical equations of the system, rather than
being related to the structure of space-time.

In the same spirit, the idea of the above mentioned authors was to explain
the multiplet structure of the genetic code through the multiplets found in
the codon representation ( = irreducible 64-dimensional representation)
of an appropriate simple Lie algebra and its branching rules into irreducible
representations of its semisimple subalgebras. They checked the tables of
branching rules of McKay and Patera \cite{mkp} for semisimple subalgebras of
simple Lie algebras of rank $\leq 8$. The most suitable multiplet structure
found is derived from the codon representation of the symplectic algebra
$\mbox{\fraktw sp}(6)$ by the following sequence of symmetry breakings:
\[
 \begin{array}{cclc}
  \mbox{\fraktw sp}(6)
  & \supset & \mbox{\fraktw sp}(4) \oplus \mbox{\fraktw su}(2) & I \\[0.5ex]
  & \supset & \mbox{\fraktw su}(2) \oplus \mbox{\fraktw su}(2)
                                   \oplus \mbox{\fraktw su}(2) & II \\[0.5ex]
  & \supset & \mbox{\fraktw su}(2) \oplus \mbox{\fraktw u}(1)
                                   \oplus \mbox{\fraktw u}(1) & III/IV/V
% & \supset & su(2) \oplus u(1) \oplus su(2) & III \\
% & \supset & su(2) \oplus u(1) \oplus u(1) & IV/V
 \end{array}
\]
The sequence of steps $I - V$ is interpreted as the evolution of the 
genetic code in the early time of organic life. 

This work, which had a strong resonance in the scientific community
\cite{mad,ste,for}, raised a lot of new interesting problems. One of
these is that the last step in the symmetry breaking is incomplete: the
lifting of degeneracy by breaking the last two $\mbox{\fraktw su}(2)$
subalgebras to $\mbox{\fraktw u}(1)$ is not followed by all codon
multiplets. Only if some of them continue to represent a single
amino acid can the actual multiplet structure of the genetic code
be obtained. This ``freezing'' had already been proposed by biologists
\cite{cri} who claimed that a completely accomplished evolution of the
genetic code should have resulted in 28 amino acids \cite{juk1} (for a
more recent review including an extensive bibliography, see \cite{juk2})
-- in perfect agreement with the mathematical model under consideration.
How\-ever the phenomenon that some of the multiplets preserve a symmetry
while it is broken in others, even though it does not contradict a purely
biological theory (in fact biologists wonder why there should be a
mathematical theory at all), is quite awkward from a mathematical
point of view. 

The basic idea behind the present project, already proposed in \cite{hor},
is to investigate the ``vicinity''  of ordinary Lie algebras, namely quantum
groups and Lie superalgebras. As it turns out, the main new problem which
appears in this context, both for quantum groups and for Lie superalgebras,
is the existence of indecomposable representations, i.e., representations
which are reducible but not fully reducible: they contain irreducible
subrepresentations but cannot be decomposed into the direct sum of
irreducible subrepresentations. As a result, the representation theory
of quantum groups and of Lie superalgebras is not developed to the same
extent as that of ordinary (reductive) Lie algebras. Therefore, the task
of performing an exhaustive search is presently not feasible: one may
at best hope for partial results. Some steps in this direction have
recently been taken by various authors \cite{fag,btj,fss1}.

\section{Basic classical Lie superalgebras}

We begin by recalling that a {\em Lie superalgebra} (LSA) is a
$\mbox{\Bbbtw Z}_2$-graded vector space
\begin{equation} \label{eq:DIRDEC1}
 \cg~=~\cg_{\bar{0}} \oplus \cg_{\bar {1}}
\end{equation}
equipped with a bilinear map $\; [.\,,. ]: \cg \times \cg \rightarrow \cg \;$
called the {\em supercommutator} which is {\em homogeneous} of degree $0$
(i.e., satisfies $~\deg([X,Y]) \, = \, \deg(X) + \deg(Y)$ \linebreak for
homogeneous $\, X,Y \in \cg$), is {\em graded antisymmetric},
\[
 [Y,X]~=~-(-1)^{\deg(X) \deg(Y)} \, [X,Y] \qquad
 \mbox{for homogeneous $X,Y \in \cg$}~,
\]
and satisfies the {\em graded Jacobi identity},
\begin{eqnarray*}
 &\begin{array}{rcl}
   (-1)^{\deg(X) \deg(Z)} \, [X,[Y,Z]] \!\!
   &\!+\!&\!\! (-1)^{\deg(X) \deg(Y)} \, [Y,[Z,X]] \\[0.5ex]
   &\!+\!&\!\! (-1)^{\deg(Y) \deg(Z)} \, [Z,[X,Y]]~=~0
  \end{array}& \\[1ex]
 &\mbox{for homogeneous $X,Y,Z \in \cg$}~.&
\end{eqnarray*}
In particular, the even part $\cg_{\bar{0}}$ of $\cg$ is an ordinary Lie
algebra and the odd part $\cg_{\bar{1}}$ of $\cg$ carries a representation
of $\cg_{\bar{0}}$, i.e., is a $\cg_{\bar{0}}$-module. In the following
we shall be dealing exclusively with finite-dimensional complex Lie
superalgebras which are {\em simple}, i.e., admit no non-trivial ideals.
Such a \LSA\ is called {\em classical} if its even part $\cg_{\bar{0}}$
is reductive, that is, if it decomposes into the direct sum of its center
and a semisimple subalgebra, or equivalently, if all representations of
$\cg_{\bar{0}}$ (in particular that on $\cg_{\bar{0}}$ itself, which is the
adjoint representation, and that on $\cg_{\bar{1}}$) are completely reducible
\cite{kac1,kac2,sch,cor}. Note that this property is not guaranteed
automatically, as it would be for ordinary semisimple Lie algebras,
according to Weyl's theorem. (However, the term ``classical'' in this
context is unfortunate because it suggests that ``classical'' for
simple Lie superalgebras bears some relation to the standard term
``classical'', in the sense of ``non-exceptional'', for simple Lie
algebras, which is not the case.) A standard argument then shows
\cite{sch,cor} that a classical \LSA\ necessarily belongs to one
of the following two types:
%\vspace{-1mm}
\begin{itemize}
 \item {\em Type~I}: \\[1ex]
       The representation of $\cg_{\bar{0}}$ on $\cg_{\bar{1}}$ is the
       direct sum of two mutually conjugate irreducible representations,
       \begin{equation} \label{eq:DIRDEC2}
        \cg_{\bar{1}}~=~\cg_1 \oplus \cg_{-1}~.
       \end{equation}
       We distinguish two subcases:
 \begin{itemize}
  \item {\em Type~I$_0$}: \\[0.5ex]
        The center $\cz_{\bar{0}}$ of $\cg_{\bar{0}}$ is trivial, i.e.,
        $\cg_{\bar{0}}$ is semisimple.
  \item {\em Type~I$_1$}: \\[0.5ex]
        The center $\cz_{\bar{0}}$ of $\cg_{\bar{0}}$ is non-trivial.
        In this case, $\cz_{\bar{0}}$ is one-dimensional and~is generated
        by an element $c$ which -- when appropriately normalized -- acts
        as the identity on $\cg_1$ and as minus the identity on $\cg_{-1}$.
 \end{itemize}
 \item {\em Type~II}: \\[1ex]
       The representation of $\cg_{\bar{0}}$ on $\cg_{\bar{1}}$ is
       irreducible. In this case, the center of $\cg_{\bar{0}}$ is
       necessarily trivial, or in other words, $\cg_{\bar{0}}$ is
       semisimple.
%\vspace{-1mm}
\end{itemize}
Another important concept for the analysis and classification of simple \LSA s
is the question whether they admit non-degenerate invariant forms. Recall that
a bilinear form $\; B: \cg \times \cg \rightarrow \mbox{\Bbbtw C} \;$ is
%($\FF$ being the ground field of $\cg$, usually either $\mbox{\Bbbtw R}$
% or $\mbox{\Bbbtw C}$)
called {\em even} if it is homogeneous of degree $0$ (i.e., satisfies
$\, B(X,Y) = 0 \,$ if $\, X \in \cg_{\bar{0}}$ and $Y \in \cg_{\bar{1}} \,$
or $\, X \in \cg_{\bar{1}}$ and $Y \in \cg_{\bar{0}}$), is called {\em
graded symmetric} if
\[
 B(Y,X)~=~(-1)^{\deg(X) \deg(Y)} \, B(X,Y) \qquad
 \mbox{for homogeneous $X,Y \in \cg$}~,
\]
and is called {\em invariant} if
\[
 B([X,Y],Z)~=~B(X,[Y,Z]) \qquad
 \mbox{for homogeneous $X,Y,Z \in \cg$}~.
\]
A simple \LSA\ is called {\em basic} if it admits an even, graded symmetric,
invariant bilinear form which is non-degenerate. Note, again, that this
property is not guaranteed automatically, as it would be for ordinary
semisimple Lie algebras, according to Cartan's criterion for semisimplicity.
In fact, it turns out that an even, graded symmetric, invariant bilinear form
on a simple \LSA\ is either non-degenerate or identically zero \cite{sch,cor}
and that, in particular, the Killing form of a simple \LSA\ defined by the
supertrace operation in the adjoint representation may vanish identically.
Moreover, there are simple \LSA s whose Killing form vanishes identically
but which are still basic because they admit some other non-degenerate,
even, graded symmetric, invariant bilinear form.

The structure theory of basic classical \LSA s is to some extent analogous
to that of ordinary semisimple Lie algebras. The first step is to choose a
Cartan subalgebra $\ch$ of $\cg$, which is by definition just a Cartan
subalgebra of its even part $\cg_{\bar{0}}$: its dimension is called
the {\em rank} of $\cg$. (If $\cg_{\bar{0}}$ has a non-trivial center
$\cz_0$ and a semisimple part $\cg_{\bar{0}}^{\rm ss}$, so that~ 
\mbox{$\cg_{\bar{0}} = \cz_{\bar{0}} \oplus \cg_{\bar{0}}^{\rm ss}$},
then~ \mbox{$\ch = \cz_{\bar{0}} \oplus \ch^{\rm ss}$}, where $\ch^{\rm ss}$
is a Cartan subalgebra of $\cg_{\bar{0}}^{\rm ss}$.) As in the case of
ordinary semisimple Lie algebras, the specific choice of Cartan subalgebra
is irrelevant, since they are all conjugate \cite{sch,cor}. This gives
rise to the {\em root system}~ \mbox{$\Delta = \Delta_0 \cup \Delta_1~$}
of $\cg$, where the set $\Delta_0$ of {\em even roots} is just the root
system of $\cg_{\bar{0}}$, as an ordinary reductive Lie algebra, and
the set $\Delta_1$ of {\em odd roots} is just the weight system of
$\cg_{\bar{1}}$, as a $\cg_{\bar{0}}$-module. Again as in the case
of ordinary semisimple Lie algebras, one associates to each root
$\, \alpha \in \Delta \,$ a unique generator $\, H_\alpha \in \ch$,
defined by
\vspace{1mm}
\[
 B(H_\alpha,H)~=~\alpha(H) \qquad \mbox{for all $H \in \ch$}~,
\]
puts
\[
 (\alpha,\beta)~=~B(H_\alpha,H_\beta) \qquad
 \mbox{for $\alpha,\beta \in \Delta$}~,
\vspace{1mm}
\]
and considers the real subspace $\ch_{\mbox{\sBbb R}}$ of $\ch$ formed by
linear combinations of the $H_\alpha$ with real coefficients. However, the
restriction of the invariant form $B$ to $\ch_{\mbox{\sBbb R}}$, which
in the case of ordinary semisimple Lie algebras is positive definite when
$B$ is chosen to be the Killing form, may now be indefinite since even in
those cases where the Killing form of $\cg$ is non-degenerate, its
restriction to the simple ideals in $\cg_{\bar{0}}$ (in most cases,
there are precisely two such simple ideals) will on one of these
simple ideals be a positive multiple of its Killing form but on
the other one be a negative multiple of its Killing form, so that
even roots $\alpha$ will satisfy either $\, (\alpha,\alpha) > 0 \,$
or $\, (\alpha,\alpha) < 0 \,$ whereas odd roots $\alpha$ will in
many cases be isotropic: $\, (\alpha,\alpha) = 0$. Even worse: when
the Killing form of $\cg$ vanishes identically, it may happen that $B$
cannot be chosen to take only real values on $\, \ch_{\mbox{\sBbb R}}
\times \ch_{\mbox{\sBbb R}}$.

This unusual kind of geometry is responsible for various complications
that arise in the next steps, which are the choice of an ordering in
$\Delta$, corresponding to the choice of a system of {\em simple roots}
$\alpha_i$ ($1 \leq i \leq r$), the definition of the {\em Cartan matrix}
$a$ and the classification of the basic classical \LSA s in terms of
{\em Kac-Dynkin diagrams}. To begin with, not all orderings are
equivalent: different choices may lead to different diagrams.
To remove this kind of ambiguity, it is convenient to restrict
the allowed orderings to a specific class, corresponding to a
{\em distinguished choice} of simple roots, characterized by the
fact that there is only one simple root which is odd, whereas the
remaining ones are even. As an example, consider the class of basic
classical \LSA s $\cg$ of type~I$_1$ (see above): here, the simple even
roots are the simple roots of $\cg_{\bar{0}}$, extended to take the value
$0$ on $c$, whereas the simple odd root is minus the highest weight of
$\cg_1$, as an irreducible $\cg_{\bar{0}}$-module, which takes the value
$1$ on $c$. In general, any such ordering gives rise to a Cartan-Weyl
decomposition
\begin{equation} \label{eq:DIRDEC3}
 \cg~=~\cn^+ \oplus \ch \oplus \cn^-~,
\end{equation} 
where $\cn^+$ and $\cn^-$ are the nilpotent subalgebras spanned by the
generators corresponding to positive and negative roots, respectively.
Combining this with the direct decomposition (\ref{eq:DIRDEC1}), one
arrives at the distinguished $\mbox{\Bbbtw Z}$-gradation of $\cg$,
\begin{equation} \label{eq:DIRDEC4}
 \begin{array}{cc}
  \cg~=~\cg_1 \oplus \cg_0 \oplus \cg_{-1}
  \quad & \quad \mbox{for type~I} \\[1ex]
  \cg~=~\cg_2 \oplus \cg_1 \oplus \cg_0 \oplus \cg_{-1} \oplus \cg_{-2}
  \quad & \quad \mbox{for type~II}
 \end{array}~,
\end{equation}
where 
\begin{equation}
 \cg_1~=~\cg_{\bar{1}} \cap \cn^+ \qquad \mbox{and} \qquad
 \cg_{-1}~=~\cg_{\bar{1}} \cap \cn^-
\end{equation}
are spanned by generators corresponding to positive and negative odd roots,
respectively, whereas
\begin{equation} 
 \cg_2~=~[\cg_1,\cg_1] \qquad \mbox{and} \qquad
 \cg_{-2}~=~[\cg_{-1},\cg_{-1}]
\end{equation}
are spanned by generators that can be written as anticommutators of these.
(Nonvanishing anticommutators of this kind exist only for basic classical
\LSA s of type~II.) The simple roots are linearly independent, and their
number $r$ is equal to the rank of $\cg$, except for the basic classical
\LSA s $\cg$ of type~I$_0$, where the simple roots are subject to one
linear relation, so their number $r$ exceeds the rank of $\cg$ by $1$.
The definition of the Cartan matrix $a$ must also be modified, due to the
possible occurence of simple odd roots of length $0$. When $\, (\alpha_i,
\alpha_i) \neq 0$, one puts
\[
 a_{ij}~=~\frac{2 (\alpha_i,\alpha_j)}{(\alpha_i,\alpha_i)}~,
\]
as usual, whereas if $\, (\alpha_i,\alpha_i) = 0$, one defines
\[
 a_{ij}~=~\frac{(\alpha_i,\alpha_j)}{(\alpha_i,\alpha_{i^\prime})}~,
\]
where $i^\prime$ is an appropriately chosen index such that $\, (\alpha_i,
\alpha_{i^\prime}) \neq 0$, whose precise definition is partly a matter of
convention. With a distinguished choice of simple roots, this can only
happen for the unique simple odd root, i.e., when $\, i = s$, and the
numbering of simple roots is then arranged in such a way that either
\mbox{$i^\prime = s+1 \,$} or$\,$ \mbox{$i^\prime = s-1$.} In this way,
$\cg$ is, up to isomorphism, determined by its Cartan matrix, being
generated by $r$ positive generators $\, e_i \in \cn^+ \,$ and $r$
negative generators  $\, f_i \in \cn^- \,$ satisfying the
supercommutation relations
\[
 [e_i,f_j]~=~\delta_{ij} h_i~~,~~[h_i,h_j]~=~0~~,~~
 [h_i,e_j]~=~a_{ij} \, e_j~~,~~[h_i,f_j]~=~- a_{ij} \, f_j 
\]
(plus Serre relations that we do not write down). Finally, the Kac-Dynkin
diagram associated with $\cg$ is drawn according to the following rules:
\begin{itemize}
 \item Simple even roots $\alpha_i$ are denoted by white blobs
       \raisebox{-0.3ex}{{\LARGE $\circ$}}, while the unique
       simple odd root $\alpha_s$ is denoted by a crossed blob
       \raisebox{0.2ex}{\boldmath $\otimes$\unboldmath} if it has zero
       length and by a black blob \raisebox{-0.3ex}{{\LARGE $\bullet$}}
       if it has non-zero length.
 \item The $j^{\rm th}$ and $k^{\rm th}$ simple root are connected
       by $\, \max \{ |a_{jk}|, |a_{kj}| \} \,$ lines, except for
       the \LSA s $D(2\,|\,1;\alpha)$, where the simple odd root is
       connected to each of the two simple even roots by a single line.
 \item When the $j^{\rm th}$ and $k^{\rm th}$ simple root are connected by
       more than a single line, an arrow is drawn pointing from the longer
       one to the shorter one.
\end{itemize}
The Kac-Dynkin diagrams of all basic classical \LSA s are listed in the
following table.

\pagebreak

\begin{table}[htbp] \label{tb:KDLSA}
 \caption{Kac-Dynkin diagrams of the basic classical Lie superalgebras}
\end{table}
\[
\begin{array}{cccc}
 \mbox{LSA} & \mbox{Type} && \mbox{Diagram} \cr
 &&& \cr
 A(m\,|\,n)~=~\mbox{\fraktw sl}(m+1\,|\,n+1) & \mbox{I}_1 &&
 \begin{picture}(160,25)
  \thicklines
  \multiput(0,0)(42,0){5}{\circle{14}}
  %\put(0,15){\makebox(0.4,0.6){1}}
  %\put(42,15){\makebox(0.4,0.6){1}}
  %\put(84,15){\makebox(0.4,0.6){1}}
  %\put(126,15){\makebox(0.4,0.6){1}}
  %\put(168,15){\makebox(0.4,0.6){1}}
  \put(79,-5){\line(1,1){10}}
  \put(79,5){\line(1,-1){10}}
  \multiput(7,0)(5,0){6}{\line(1,0){3}}
  \put(49,0){\line(1,0){28}}
  \put(91,0){\line(1,0){28}}
  \multiput(133,0)(5,0){6}{\line(1,0){3}}
  \put(0,-10){$\underbrace{~~~~~~~~~~}_{m}$}
  \put(126,-10){$\underbrace{~~~~~~~~~~}_{n}$}
 \end{picture} \cr
 (m > n \geq 0) &&&\cr&&&\cr
 A(n\,|\,n)~=~\mbox{\fraktw sl}(n+1\,|\,n+1)/\langle 1 \rangle & \mbox{I}_0 &&
 \begin{picture}(160,25)
  \thicklines
  \multiput(0,0)(42,0){5}{\circle{14}}
  %\put(0,15){\makebox(0.4,0.6){1}}
  %\put(42,15){\makebox(0.4,0.6){1}}
  %\put(84,15){\makebox(0.4,0.6){1}}
  %\put(126,15){\makebox(0.4,0.6){1}}
  %\put(168,15){\makebox(0.4,0.6){1}}
  \put(79,-5){\line(1,1){10}}
  \put(79,5){\line(1,-1){10}}
  \multiput(7,0)(5,0){6}{\line(1,0){3}}
  \put(49,0){\line(1,0){28}}
  \put(91,0){\line(1,0){28}}
  \multiput(133,0)(5,0){6}{\line(1,0){3}}
  \put(0,-10){$\underbrace{~~~~~~~~~~}_{n}$}
  \put(126,-10){$\underbrace{~~~~~~~~~~}_{n}$}
 \end{picture} \cr
 (n \geq 1) &&&\cr&&&\cr
 B(m\,|\,n) = \mbox{\fraktw osp}(2m+1\,|\,2n) & \mbox{II} &&
 \begin{picture}(220,25)
 %\begin{picture}(200,25)
  \thicklines
  \multiput(0,0)(42,0){6}{\circle{14}}
  %\put(0,15){\makebox(0.4,0.6){2}}
  %\put(42,15){\makebox(0.4,0.6){2}}
  %\put(84,15){\makebox(0.4,0.6){2}}
  %\put(126,15){\makebox(0.4,0.6){2}}
  %\put(168,15){\makebox(0.4,0.6){2}}
  %\put(210,15){\makebox(0.4,0.6){2}}
  \put(79,-5){\line(1,1){10}}
  \put(79,5){\line(1,-1){10}}
  \multiput(7,0)(5,0){6}{\line(1,0){3}}
  \put(49,0){\line(1,0){28}}
  \put(91,0){\line(1,0){28}}
  \multiput(133,0)(5,0){6}{\line(1,0){3}}
  \put(174,-3){\line(1,0){30}}
  \put(174,3){\line(1,0){30}}
  \put(195,0){\line(-1,1){10}}\put(195,0){\line(-1,-1){10}}
  \put(0,-10){$\underbrace{~~~~~~~~~~}_{n-1}$}
  \put(126,-10){$\underbrace{~~~~~~~~~~}_{m-1}$}
 \end{picture} \cr
 (m,n \geq 1) &&&\cr&&&\cr
 B(0\,|\,n) = \mbox{\fraktw osp}(1\,|\,2n) & \mbox{II} &&
 \begin{picture}(80,25)
  \thicklines
  \put(0,0){\circle{14}}
  \put(42,0){\circle{14}}
  \put(84,0){\circle*{14}}
  %\put(0,15){\makebox(0.4,0.6){2}}
  %\put(42,15){\makebox(0.4,0.6){2}}
  %\put(84,15){\makebox(0.4,0.6){2}}
  \multiput(7,0)(5,0){6}{\line(1,0){3}}
  \put(48,-3){\line(1,0){30}}
  \put(48,3){\line(1,0){30}}
  \put(69,0){\line(-1,1){10}}
  \put(69,0){\line(-1,-1){10}}
  \put(0,-10){$\underbrace{~~~~~~~~~~}_{n-1}$}
 \end{picture} \cr
 (n \geq 1) &&&\cr&&&\cr
 C(n+1) = \mbox{\fraktw osp}(2\,|\,2n) & \mbox{I}_1 &&
 \begin{picture}(120,25)
  \thicklines
  \multiput(0,0)(42,0){4}{\circle{14}}
  %\put(0,15){\makebox(0.4,0.6){1}}
  %\put(42,15){\makebox(0.4,0.6){2}}
  %\put(84,15){\makebox(0.4,0.6){2}}
  %\put(126,15){\makebox(0.4,0.6){1}}
  \put(-5,-5){\line(1,1){10}}
  \put(-5,5){\line(1,-1){10}}
  \put(7,0){\line(1,0){28}}
  \multiput(49,0)(5,0){6}{\line(1,0){3}}
  \put(101,0){\line(1,1){10}}
  \put(101,0){\line(1,-1){10}}
  \put(90,-3){\line(1,0){30}}
  \put(90,3){\line(1,0){30}}
  \put(42,-10){$\underbrace{~~~~~~~~~~}_{n-1}$}
 \end{picture} \cr
 (n \geq 1) &&&\cr &&&\cr
 D(m\,|\,n) = \mbox{\fraktw osp}(2m\,|\,2n) & \mbox{II} &&
 \begin{picture}(220,25)
 %\begin{picture}(200,25)
  \thicklines
  \multiput(0,0)(42,0){5}{\circle{14}}
  %\put(0,15){\makebox(0.4,0.6){2}}
  %\put(42,15){\makebox(0.4,0.6){2}}
  %\put(84,15){\makebox(0.4,0.6){2}}
  %\put(126,15){\makebox(0.4,0.6){2}}
  %\put(168,15){\makebox(0.4,0.6){2}}
  \put(79,-5){\line(1,1){10}}
  \put(79,5){\line(1,-1){10}}
  \multiput(7,0)(5,0){6}{\line(1,0){3}}
  \put(49,0){\line(1,0){28}}
  \put(91,0){\line(1,0){28}}
  \multiput(133,0)(5,0){6}{\line(1,0){3}}
  \put(173,5){\line(2,1){20}}
  \put(173,-5){\line(2,-1){20}}
  \put(200,19){\circle{14}}
  %\put(215,20){\makebox(0.4,0.6){1}}
  \put(200,-19){\circle{14}}
  %\put(215,-20){\makebox(0.4,0.6){1}}
  \put(0,-10){$\underbrace{~~~~~~~~~~}_{n-1}$}
  \put(126,-10){$\underbrace{~~~~~~~~~~}_{m-2}$}
 \end{picture} \cr
 (m \geq 3, n \geq 1) &&&\cr&&&\cr
 D(2\,|\,n) = \mbox{\fraktw osp}(4\,|\,2n) & \mbox{II} &&
 \begin{picture}(120,25)
  \thicklines
  \multiput(0,0)(42,0){3}{\circle{14}}
  %\put(0,15){\makebox(0.4,0.6){2}}
  %\put(42,15){\makebox(0.4,0.6){2}}
  %\put(84,15){\makebox(0.4,0.6){2}}
  \put(79,-5){\line(1,1){10}}
  \put(79,5){\line(1,-1){10}}
  \multiput(7,0)(5,0){6}{\line(1,0){3}}
  \put(49,0){\line(1,0){28}}
  \put(91,5){\line(2,1){20}}
  \put(91,-5){\line(2,-1){20}}
  \put(117,19){\circle{14}}
  %\put(132,20){\makebox(0.4,0.6){1}}
  \put(117,-19){\circle{14}}
  %\put(132,-20){\makebox(0.4,0.6){1}}
  \put(0,-10){$\underbrace{~~~~~~~~~~}_{n-1}$}
 \end{picture} \cr
 (n \geq 1) &&&\cr&&&\cr
 D(2\,|\,1;\alpha) = \mbox{\fraktw osp}(4\,|\,2\,;\alpha) & \mbox{II} &&
 \begin{picture}(60,25)
  \thicklines
  \put(10,0){\circle{14}}
  %\put(-10,0){\makebox(0.4,0.6){2}}
  \put(5,-5){\line(1,1){10}}
  \put(5,5){\line(1,-1){10}}
  \put(15,5){\line(2,1){20}}
  \put(15,-5){\line(2,-1){20}}
  \put(42,19){\circle{14}}
  %\put(57,20){\makebox(0.4,0.6){1}}
  \put(42,-19){\circle{14}}
  %\put(57,-20){\makebox(0.4,0.6){1}}
 \end{picture} \cr
 (\alpha \neq 0,-1,\infty) &&&\cr&&&\cr
 F(4) & \mbox{II} &&
 \begin{picture}(120,25)
  \thicklines
  \multiput(0,0)(42,0){4}{\circle{14}}
%  \put(0,0){\circle{14}}
%  \put(42,0){\circle{14}}
%  \put(84,0){\circle{14}}
%  \put(126,0){\circle{14}}
  %\put(0,15){\makebox(0.4,0.6){2}}
  %\put(42,15){\makebox(0.4,0.6){3}}
  %\put(84,15){\makebox(0.4,0.6){2}}
  %\put(126,15){\makebox(0.4,0.6){1}}
  \put(-5,-5){\line(1,1){10}}
  \put(-5,5){\line(1,-1){10}}
  \put(7,0){\line(1,0){28}}
  \put(48,-3){\line(1,0){30}}
  \put(48,3){\line(1,0){30}}
  \put(59,0){\line(1,1){10}}
  \put(59,0){\line(1,-1){10}}
  \put(91,0){\line(1,0){28}}
 \end{picture} \cr
 &&&\cr&&&\cr
 G(3) & \mbox{II} &&
 \begin{picture}(80,25)
  \thicklines
  \multiput(0,0)(42,0){3}{\circle{14}}
%  \put(0,0){\circle{14}}
%  \put(42,0){\circle{14}}
%  \put(84,0){\circle{14}}
  %\put(0,15){\makebox(0.4,0.6){2}}
  %\put(42,15){\makebox(0.4,0.6){4}}
  %\put(84,15){\makebox(0.4,0.6){2}}
  \put(-5,-5){\line(1,1){10}}
  \put(-5,5){\line(1,-1){10}}
  \put(7,0){\line(1,0){28}}
  \put(48,-4){\line(1,0){30}}
  \put(49,0){\line(1,0){28}}
  \put(48,4){\line(1,0){30}}
  \put(59,0){\line(1,1){10}}
  \put(59,0){\line(1,-1){10}}
 \end{picture}
\end{array}
\]

\noindent
Observe that the Cartan matrix cannot always be reconstructed uniquely
from the corresponding Kac-Dynkin diagram, in particular this happens
for the \LSA s $D(2\,|\,1;\alpha)$.

The basic classical \LSA s of type~I$_0$ are in many respects patho\-logical,
but almost all the general results about basic classical Lie superalgebras
\linebreak (including the main ones from representation theory) remain true
if one replaces \linebreak $A(n\,|\,n) = \mbox{\fraktw sl}(n+1\,|\,n+1)/
\langle 1 \rangle~$ by its natural central extension $\, \mbox{\fraktw sl}%
(n+1\,|\,n+1)$. (This leads, for example, to an enrichment of the
representation theory, since the irreducible representations of the
former form a subclass of the irreducible representations of the latter:
namely those in which the central element is represented by the zero
operator.) We shall therefore, throughout the rest of this paper,
adopt the following terminology:
\begin{itemize}
 \item Type~I \LSA s: \\[1ex]
       $\mbox{\fraktw sl}(p\,|\,q) \,$ with $\, p \geq q \geq 1 \,$ and
       $\, (p,q) \neq (1,1) \,$ \\
       (the case $\, p = q =1 \,$ is excluded since $A(0,0)$ is not simple),
       \\[0.5ex]
       $\mbox{\fraktw osp}(2\,|\,2n) \,$ with $\, n \geq 1 \,$
       and $\, n \neq 1 \,$ \\
       (the case $\, n =1 \,$ is excluded since
       $\, \mbox{\fraktw osp}(2\,|\,2) \cong \mbox{\fraktw sl}(3\,|\,2)$).
 \item Type~II \LSA s: \\[1ex]
       $\mbox{\fraktw osp}(p\,|\,2n) \,$ with $\, p = 1 \,$ or $\, p \geq 3$
       and $n \neq 1$,
       \\[0.5ex]
       $\mbox{\fraktw osp}(4\,|\,2\,;\alpha)$, $F(4)$, $G(3)$.
\end{itemize}
It is also interesting to compare the Kac-Dynkin diagram of $\cg$
with the Dynkin diagram of its even part $\cg_{\bar{0}}$ and the Dynkin
diagram of the subalgebra $\cg_0$ that appears in the direct decomposition
(\ref{eq:DIRDEC3}). For type~I \LSA s, where \mbox{$\cg_{\bar{0}} = \cg_0$,}
the latter is obtained from the former by simply removing the simple odd
root $\alpha_s$, which may therefore be thought of as representing the
one-dimensional center of the even part, whereas for type~II \LSA s, the
Dynkin diagram of $\cg_0$ is obtained from the Kac-Dynkin diagram of $\cg$
by removing the simple odd root $\alpha_s$ and from the Dynkin diagram of
$\cg_{\bar{0}}$ by removing one of its simple roots: this simple root,
which we shall denote by $\alpha_s^0$, is usually referred to as the
``hidden'' simple root of $\cg_{\bar{0}}$ because in the Kac-Dynkin
diagram of $\cg$, it can be thought of as being ``hidden behind'' the
simple odd root $\alpha_s$.

\section{Representation theory}

The representation theory of basic classical \LSA s $\cg$ (with 
$\; A(n\,|\,n) = \mbox{\fraktw sl}(n+1\,|\,n+1)/\langle 1 \rangle \;$
replaced by $\, \mbox{\fraktw sl}(n+1\,|\,n+1)$; see above) has been
developed by Kac \cite{kac1,kac2}. Using the Poincar\'e-Birkhoff-Witt
theorem, the finite-dimensional irreducible representations of $\cg$
are constructed by the method of induced representations, that is, as
quotient spaces of Verma modules by their maximal invariant subspaces.
This implies that all finite-dimensional irreducible representations of
$\cg$ are highest weight representations, that is, representations of the
form~$\pi_\Lambda : \cg \rightarrow \mbox{End}(V_\Lambda)~$ associated
to a highest weight $\, \Lambda \in \ch^* \,$ and characterized by the
presence of a non-zero cyclic vector $\, v_\Lambda \in V_\Lambda \,$
satisfying
\[
 \cn^+(v_\Lambda)~=~0 \qquad \mbox{and} \qquad
 H(v_\Lambda)~=~\Lambda(H) \, v_\Lambda \quad \mbox{for all $\, H \in \ch$}~.
\]
A necessary condition for such a representation to be finite-dimensional
is that $\Lambda$ is {\em dominant integral}, which means that the Dynkin
labels
\begin{equation} \label{eq:DYNL1}
 l_i~=~\frac{2 (\Lambda,\alpha_i)}{(\alpha_i,\alpha_i)}~
\end{equation}
of $\Lambda$ associated with the simple even roots $\alpha_i$
($i = 1,\ldots,r$, $i \neq s$) of $\cg$ must be non-negative
integers. For type~I \LSA s, this is the only condition to
be imposed. In particular, the value of the Dynkin label
\begin{equation} \label{eq:DYNL2}
 l_s~=~\frac{(\Lambda,\alpha_s)}{(\alpha_s,\alpha_{s^\prime})}~
\end{equation}
of $\Lambda$ associated with the simple odd root $\alpha_s$ of $\cg$ may
in this case be an arbitrary complex number, whereas for type~II \LSA s,
it is subject to additional restrictions: some of these simply express
the requirement that the Dynkin label
\begin{equation} \label{eq:DYNL3}
 l_s^{\,0}~=~\frac{2 (\Lambda,\alpha_s^0)}{(\alpha_s^0,\alpha_s^0)}~
\end{equation}
of $\Lambda$ associated with the hidden simple root $\alpha_s^0$
of $\cg_{\bar{0}}$ must also be a non-negative integer, while the
others are supplementary conditions to guarantee that $\Lambda$ is
the highest weight of a finite-dimensional irreducible representation
not only of $\cg_{\bar{0}}$ but also of $\cg$. For detailed formulae,
see \cite{cor,kac1,kac2,fss2}.

An explicit construction of the representation~
\mbox{$\pi_\Lambda: \cg \rightarrow \mbox{End}(V_\Lambda)~$}
of $\cg$ starts out from the representation~
$\pi_{\Lambda,0}: \cg_0 \rightarrow \mbox{End}(V_{\Lambda,0})~$
of $\cg_0$ with highest weight $\Lambda$, or more precisely, with
highest weight given by the restriction of $\Lambda$ to the intersection
of $\cg_0$ with the Cartan subalgebra $\ch$ of $\cg$. This representation
is first extended to a representation of the subalgebra $\; \ck = \cg_0
\oplus \cg_1 \oplus \cg_2 \;$ by letting $\, \cg_1 \oplus \cg_2 \,$ act
trivially on $V_{\Lambda,0}$. Then define
\begin{equation} \label{INDMOD}
 \begin{array}{cc} 
  \bar{V}_\Lambda~=~\mbox{Ind}^{\cg}_{\ck} \, V_{\Lambda,0}
  \quad & \quad \mbox{for type~I \LSA s}~, \\[1ex]
  \bar{V}_\Lambda~=~\mbox{Ind}^{\cg}_{\ck} \, V_{\Lambda,0} \, / \, M
  \quad & \quad \mbox{for type~II \LSA s}~,
 \end{array}
\end{equation} 
where the invariant submodule $M$ is obtained by applying arbitrary linear
combinations of products of elements of $\cg$ (i.e., the enveloping algebra
$U(\cg)$) to the vector obtained by $(l_s^{\,0}+1)$-fold application of the
even generator $\, E_{-\alpha_s^0} \in \cg_{-2} \,$ to the highest weight
vector $v_\Lambda$:
\[
 M~=~\langle \, U(\cg) \, E_{-\alpha_s^0}^{l_s^{\,0}+1} v_\Lambda \rangle~.
\]
The Kac module $\bar{V}_\Lambda$ is finite-dimensional and contains
a unique maximal submodule $\bar{I}_\Lambda$. Then
\begin{equation}
 V_\Lambda~=~\bar{V}_\Lambda \, / \, \bar{I}_\Lambda~.
\end{equation}
Any finite-dimensional irreducible representation of $\cg$ can be obtained
in this way. However, it is in general difficult to gain control over the
submodule $\bar{I}_\Lambda$, so explicit calculations are usually only
possible when this submodule vanishes~-- which is one of the main reasons
for the special role played by the so-called {\em typical representations}:
\[
 \bar{I}_\Lambda~=~\{0\}~~,~~V_\Lambda~=~\bar{V}_\Lambda \qquad
 \mbox{for typical representations}
\]
Typical representations are, by definition, irreducible representations
that may appear as direct summands in completely reducible representations
only, whereas irreducible representations appearing as subrepresentations of
indecomposable \linebreak (that is, reducible but not completely reducible)
representations are called \linebreak atypical. A useful criterion
for an irreducible representation to be typical is that \linebreak
\mbox{$(\Lambda + \rho,\alpha) \neq 0$}~ for all odd roots $\alpha$
for which $2\alpha$ is not an even root, where
\[
 \rho~=~\rho_0 \, - \, \rho_1~~,~~
 \rho_0~=~{\textstyle{1 \over 2}} \sum_{\alpha \in \Delta_0^+} \alpha~~,~~
 \rho_1~=~{\textstyle{1 \over 2}} \sum_{\alpha \in \Delta_1^+} \alpha
\] 
Denoting the number of positive odd roots, i.e., the cardinality of
$\Delta_1^+$, by $N_1$ (and similarly, the number of positive even roots,
i.e., the cardinality of $\Delta_0^+$, by $N_0$), one can write down an
explicit formula for the total dimension of any typical representation:
\begin{equation} \label{eq:DIMFOR1}
 \dim V_\Lambda~=~2^{N_1} \, \prod_{\alpha \in \Delta_0^+} 
                  \frac{(\Lambda + \rho,\alpha)}{(\rho_0,\alpha)}~.
\end{equation}
This formula can be simplified by expressing the product on the rhs in
terms \linebreak of the standard Weyl dimension formula for an irreducible
representation \linebreak \mbox{$\pi_{\tilde{\Lambda},\bar{0}}: \cg_{\bar{0}}
\rightarrow \mbox{End}(V_{\tilde{\Lambda},\bar{0}})~$} of the even part
$\cg_{\bar{0}}$ of $\cg$ with highest weight $\tilde{\Lambda}$:
\begin{equation} \label{eq:DIMFOR2}
 \dim V_{\tilde{\Lambda},\bar{0}}~
 =~\prod_{\alpha \in \Delta_0^+} 
   \frac{(\tilde{\Lambda} + \rho_0,\alpha)}{(\rho_0,\alpha)}~.
\end{equation}
To establish the desired relation, observe that $\, (\rho_1,\alpha_i)
= 0 \,$ for all simple even roots $\alpha_i$ of $\cg$ because the
corresponding positive and negative root generators $E_{\alpha_i}$
and $E_{-\alpha_i}$ belong to $\cg_0$ and hence preserve the
subspaces in the direct decomposition (\ref{eq:DIRDEC4}):
therefore, the number $(2\rho_1,\alpha_i)$, which is precisely
the trace of the operator on $\cg_1$ that represents$\,$
\mbox{$H_{\alpha_i} = [E_{\alpha_i},E_{-\alpha_i}]$,} must vanish.
Therefore, for type~I \LSA s, we may simply put $\, \tilde{\Lambda}
= \Lambda$, so
\begin{equation} \label{eq:DIMFOR3}
 \dim V_\Lambda~=~2^{N_1} \, \dim V_{\Lambda,\bar{0}}~.
\end{equation}
An alternative argument for deriving this formula is to use the construction
of the Kac module because, in this case, \mbox{$\cg_{\bar{0}} = \cg_0$,}
\mbox{$V_{\Lambda,\bar{0}} = V_{\Lambda,0} \,$} and$\,$ \mbox{$[E_\alpha,
E_\beta] = 0$} for all positive odd roots$\,$ \mbox{$\alpha,\beta \in
\Delta_1^+$,} so that
\[
 V_\Lambda~=~\bar{V}_\Lambda~=~\mbox{Ind}^{\cg}_{\ck} \, V_{\Lambda,0}~
 \cong~\Wedge \cg_{-1} \otimes V_{\Lambda,0}~,
\]
where $\Wedge \cg_{-1}$ denotes the exterior or Grassmann algebra over
$\cg_{-1}$, which has dimension $2^{N_1}$. For type~II \LSA s, we let
$\{\lambda_1,\ldots,\lambda_{s-1},\lambda_s^0,\lambda_{s+1},\ldots,
\lambda_r\}$ denote the basis of fundamental weights dual to the basis
$\{\alpha_1,\ldots,\alpha_{s-1},\alpha_s^0, \linebreak \alpha_{s+1},
\ldots,\alpha_r\}$ of simple roots for $\cg_{\bar{0}}$ and introduce
the shifted highest weight
\begin{equation} \label{eq:DIMFOR4}
 \tilde{\Lambda}~=~\Lambda \, - \,
                   {2 (\rho_1,\alpha_s^0) \over (\alpha_s^0,\alpha_s^0)} \,
                   \lambda_s^0~,
\end{equation}
which in terms of Dynkin labels means
\begin{equation} \label{eq:DIMFOR5}
 \tilde{l_i} = l_i \quad \mbox{for} \quad i \neq s~~,~~
 \tilde{l}_s^{\,0}~=~l_s^{\,0} \, - \,
                     {2 (\rho_1,\alpha_s^0) \over (\alpha_s^0,\alpha_s^0)}~.
\end{equation}
It should be noted that although the original highest weight $\Lambda$
is dominant integral, the shifted highest weight $\tilde{\Lambda}$ need
not be, since $\, 2 (\rho_1,\alpha_s^0) / (\alpha_s^0,\alpha_s^0) \,$ may
assume half-integer values (see Table 2), so $\tilde{l}_s^{\,0}$ may become
half-integer and/or negative. In this case, equation (\ref{eq:DIMFOR2})
is only formal, in the sense that the expression ``$\dim V_{\tilde{\Lambda},
\bar{0}}$'' does not necessarily stand for the dimension of an irreducible
representation of $\cg_{\bar{0}}$. Therefore, we introduce for every
ordinary semisimple Lie algebra $\ca$ of rank $p$ the abbreviation
$d_{\mbox{\sfrak a}}$ to denote the dimension function for its irreducible
representations, which is a polynomial in $p$ variables given by the standard
Weyl dimension formula, and we simply write $d_{\bar{0}}$ instead of
$d_{\mbox{\sfrak g}_{\bar{0}}}$, so equation (\ref{eq:DIMFOR2}) is
replaced by
\begin{equation} \label{eq:DIMFOR6}
 d_{\bar{0}}(\tilde{\Lambda})~
 =~\prod_{\alpha \in \Delta_0^+} 
   \frac{(\tilde{\Lambda} + \rho_0,\alpha)}{(\rho_0,\alpha)}~.
\end{equation}
Then equation (\ref{eq:DIMFOR1}) becomes
\begin{equation} \label{eq:DIMFOR7}
 \dim V_\Lambda~=~2^{N_1} \, d_{\bar{0}}(\tilde{\Lambda})~.
\vspace{1mm}
\end{equation}
In order to proceed further, we need more information on the behavior
of the function $d_{\bar{0}}$. First of all, we observe that as long as
\begin{equation} \label{eq:POSITW1}
 l_s^{\,0} \geq b \qquad \mbox{i.e.} \qquad
 \tilde{l}_s^{\,0}~\geq~- \, {\textstyle {1 \over 2}}~,
\end{equation}
where $b$ is the integer part of $\, 2 (\rho_1,\alpha_s^0) /
(\alpha_s^0,\alpha_s^0) \,$ (see Table 2), all factors in the
product on the rhs of equation (\ref{eq:DIMFOR6}) remain positive.
\begin{table}[ht] \label{tb:SHW}
 \caption{Shift of highest weight for type II Lie superalgebras}
 \vspace{3ex}
 \[ \mbox{
 \begin{tabular}{|c|c|c|} \hline
  \rule[-3ex]{0ex}{7.5ex} LSA &
  ${\displaystyle {2 (\rho_1,\alpha_s^0) \over (\alpha_s^0,\alpha_s^0)}}$ & $b$
  \\ \hline\hline
  \rule{0ex}{3ex}
  $B(m\,|\,n) = \mbox{\fraktw osp}(2m+1\,|\,2n)$ & & \\
  \rule[-1.5ex]{0ex}{1.5ex} $(m,n \geq 1)$ &
  \raisebox{1.5ex}[-1.5ex]{$m + {1 \over 2}$} &
  \raisebox{1.5ex}[-1.5ex]{$m$} \\ \hline
  \rule{0ex}{3ex}
  $B(0\,|\,n) = \mbox{\fraktw osp}(1\,|\,2n)$ & & \\
  \rule[-1.5ex]{0ex}{1.5ex} $(n \geq 1)$ &
  \raisebox{1.5ex}[-1.5ex]{${1 \over 2}$} &
  \raisebox{1.5ex}[-1.5ex]{$0$} \\ \hline
  \rule{0ex}{3ex}
  $D(m\,|\,n) = \mbox{\fraktw osp}(2m\,|\,2n)$ & & \\
  \rule[-1.5ex]{0ex}{1.5ex} $(m \geq 2, n \geq 1)$ &
  \raisebox{1.5ex}[-1.5ex]{$m$} &
  \raisebox{1.5ex}[-1.5ex]{$m$} \\ \hline
  \rule{0ex}{3ex}
  $D(2\,|\,1;\alpha) = \mbox{\fraktw osp}(4\,|\,2\,;\alpha)$ & & \\
  \rule[-1.5ex]{0ex}{1.5ex} $(\alpha \neq 0,-1,\infty)$ &
  \raisebox{1.5ex}[-1.5ex]{$2$} &
  \raisebox{1.5ex}[-1.5ex]{$2$} \\ \hline
  \rule[-1.5ex]{0ex}{4.5ex} $F(4)$ &      $4$      & $4$ \\ \hline
  \rule[-2ex]{0ex}{5.5ex}   $G(3)$ & ${7 \over 2}$ & $3$ \\ \hline
 \end{tabular}
 } \]
\end{table}
Hence in this region, $d_{\bar{0}}$ is positive and monotonically
increasing in the following sense: Suppose that $\tilde{\Lambda}$
and $\tilde{M}$ are two highest weights for $\cg_{\bar{0}}$, with
Dynkin labels $l_i$, $\tilde{l}_s^{\,0}$ and $m_i$, $\tilde{m}_s^0$,
respectively, where $\, i = 1,\ldots,r,~i \neq s \,$ and
$\, \tilde{l}_s^{\,0}, \tilde{m}_s^0 \geq - {1 \over 2}$.
Then defining
\begin{equation} \label{eq:MONOTO1}
 \tilde{\Lambda} \geq \tilde{M} \qquad \Longleftrightarrow \qquad
 l_i \geq m_i~~(1 \leq i \leq r,~i \neq s)~~\mbox{and}~~
 \tilde{l}_s^{\,0} \geq \tilde{m}_s^0~,
\end{equation}
and $\, \tilde{\Lambda} > \tilde{M} \,$ iff $\, \tilde{\Lambda} \geq
\tilde{M} \,$ and $\, \tilde{\Lambda} \neq \tilde{M}$, we have
\begin{equation} \label{eq:MONOTO2}
 \begin{array}{ccc}
  \tilde{\Lambda} \geq \tilde{M} \quad &\Longrightarrow& \quad
  d_{\bar{0}}(\tilde{\Lambda}) \geq d_{\bar{0}}(\tilde{M})~, \\[1mm]
  \tilde{\Lambda}  >   \tilde{M} \quad &\Longrightarrow& \quad
  d_{\bar{0}}(\tilde{\Lambda})  >   d_{\bar{0}}(\tilde{M})~.
 \end{array}
\end{equation}
Another important observation is that when the inequality (\ref{eq:POSITW1})
does not hold, then the Dynkin labels $\, l_1 , \ldots , l_r \,$ of $\Lambda$
must satisfy certain supplementary conditions which can be shown to imply that
the representation of $\cg$ characterized by the highest weight $\Lambda$ is
atypical; see below. As we are only interested in typical representations,
we may therefore impose the inequality (\ref{eq:POSITW1}) and make use of
the monotonicity property (\ref{eq:MONOTO2}) to provide lower bounds for
the expression in equation (\ref{eq:DIMFOR7}). There is also an abstract
argument to show that the function $d_{\bar{0}}$ continues to take integer
values as long as $\, 2 (\rho_1,\alpha_s^0) / (\alpha_s^0,\alpha_s^0) \,$
is an integer, due to the following

\noindent
{\em Proposition:}~ Let $P$ be a polynomial of degree $r$ in one real variable
which takes integer values on all integers greater than some fixed integer.
Then $P$ takes integer values on all integers.

\vspace{-1mm}
\small
\begin{quote}
 {\em Proof:}~The basic trick for the proof is to expand the polynomial $P$
 not in the standard basis of polynomials $x^l$ $(l = 0,1,\ldots,r)$ but in
 a different basis of polynomials defined by the binomial coefficients, that
 is, to write
 \begin{equation} \label{eq:BINEXP1}
  P(x)~=~\sum_{l=0}^r \, a_l \, {x \choose l}~
       =~\sum_{l=0}^r \, {a_l \over l!} \; x \, (x-1) \ldots (x-l+1)~.
 \end{equation}
 Observing that
 $$
  {x+1 \choose l} - {x \choose l}~=~{x \choose l-1}
 $$
 and therefore
 $$
  P(x+1) \, - P(x)~=~\sum_{k=0}^{r-1} \, a_{k+1} \, {x \choose k}~,
 $$
 we may conclude by induction on $r$ that the property of $P(n)$ being
 an integer for all $\, n \smin \mbox{\Bbbtw Z} \,$ and the -- apparently
 weaker -- property of $P(n)$ being an integer for all $\, n \smin
 \mbox{\Bbbtw Z} \,$ satisfying $\, n \geq n_0 \,$ for some $\, n_0
 \smin \mbox{\Bbbtw Z} \,$ are both equivalent to the fact that the
 coefficients $a_l$ of $P$ in the expansion (\ref{eq:BINEXP1}) are
 all integers; in fact, they can be computed recursively from the
 formula
 \begin{equation} \label{eq:BINEXP2}
  \sum_{i=0}^p \, (-1)^{p-i} \, {p \choose i} \, P(x+i)~
  =~\sum_{k=0}^{r-p} \, a_{k+p} \, {x \choose k}~,
 \end{equation}
 which in turn can be inferred from the previous one by induction on $p$.
\vspace{-1mm}
\end{quote}
\normalsize

\noindent
According to Table 2, this implies that the only type~II \LSA s for which
$d_{\bar{0}}$ may take non-integer values and hence $\dim V_\Lambda$ need
no longer be a multiple of~$2^{N_1}$ are those belonging to the series
$\, B(m\,|\,n) = \mbox{\fraktw osp}(2m+1\,|\,2n)$ $(m,n \geq 1)$, those
belonging to the series $\, B(0\,|\,n) = \mbox{\fraktw osp}(1\,|\,2n)$
$(n \geq 1) \,$ and, finally, the exceptional \LSA\ $G(3)$.

With these generalities out of the way, we can proceed to determine the
typical codon representations, that is, the $64$-dimensional irreducible
representations, of basic classical \LSA s. For type~I \LSA s, this is
easily done by exploiting the dimension formula (\ref{eq:DIMFOR3}), which
implies that the number $N_1$ of positive odd roots must not exceed $6$ and
that $\Lambda$ must be the highest weight of an irreducible representation
of $\cg_{\bar{0}}$ of dimension $2^{\,6-N_1}$:
\begin{itemize}
 \item The series $\, \mbox{\fraktw sl}(m+1\,|\,n+1) \,$
       with $\, m > n \geq 0 \,$: \\[1mm]
       Here, $N_1$ equals $\, (m+1)(n+1)$, so we must have $\, m \leq 2$,
       $n \leq 1$, which leaves the following possibilities: \\[1mm]
       either $\, n = 0 \,$ and $\, m = 0,1,2,3,4,5$, \\[1mm]
       or $\, n = 1 \,$ and $\, m = 2$.
 \item The series $\, \mbox{\fraktw sl}(n+1\,|\,n+1) \,$
       with $\, n \geq 1 \,$: \\[1mm]
       Here, $N_1$ equals $\, (n+1)^2$, so we must have $\, n = 1$.
 \item The series $\, \mbox{\fraktw osp}(2\,|\,2n) \,$
       with $\, n \geq 2 \,$: \\[1mm]
       Here, $N_1$ equals $\, 2n$, so we must have $\, n \leq 3$.

\end{itemize}
This leads to the list of typical codon representations of type~I \LSA s
presented in Table 3. Note that the coefficient $l_s$ of $\Lambda$ along
the simple odd root $\alpha_s$ remains unspecified: it can take any complex
value except $0$ and a few other integers that must be excluded in order to
guarantee that the representation is indeed typical, and its choice has no
influence on the dimension of the representation.

\vfill

\begin{table}[ht] \label{tb:CRTYPE1}
 \caption{Typical codon representations of type~I \LSA s}
 \vspace{3ex}
 \[ \mbox{
 \begin{tabular}{|c|c|c|c|c|} \hline
  \rule{0ex}{3ex}           Lie            &                                 &
                            Highest Weight & Highest Weight     & Typicality \\
  \rule[-1.5ex]{0ex}{1.5ex} Superalgebra   & \raisebox{1.5ex}[-1.5ex]{$N_1$} &
                            of $\cg$       & of $\cg_{\bar{0}}$ & Condition
  \\ \hline\hline
  \rule[-1.5ex]{0ex}{4.5ex} $\mbox{\fraktw sl}(2\,|\,1)$      & $2$ &
                            $(15,l_2)$        &   $15$        &
                            $- l_2 \neq 0,16$        \\ \hline
  \rule[-1.5ex]{0ex}{4.5ex} $\mbox{\fraktw sl}(3\,|\,1)$      & $3$ &
                            $(1,1,l_3)$       & $(1,1)$       &
                            $- l_3 \neq 0,2,4$       \\ \hline
  \rule{0ex}{3ex}                                             &     &
                            $(1,0,0,l_4)$     & $(1,0,0)$     &
                            $- l_4 \neq 0,1,2,4$     \\
  \rule[-1.5ex]{0ex}{1.5ex}
   \raisebox{1.5ex}[-1.5ex]{$\mbox{\fraktw sl}(4\,|\,1)$}     &
   \raisebox{1.5ex}[-1.5ex]{$4$} &
                            $(0,0,1,l_4)$     & $(0,0,1)$     &
                            $- l_4 \neq 0,2,3,4$     \\ \hline
  \rule[-1.5ex]{0ex}{4.5ex} $\mbox{\fraktw sl}(6\,|\,1)$      & $6$ &
                            $(0,0,0,0,0,l_6)$ & $(0,0,0,0,0)$ &
                            $- l_6 \neq 0,1,2,3,4,5$ \\ \hline
  \rule{0ex}{3ex}                                             &     &
                            $(3,l_2,0)$       & $(3)-(0)$     &
                            $l_2 \neq -4,-3,0,1$     \\
                            $\mbox{\fraktw sl}(2\,|\,2)$      & $4$ &
                            $(1,l_2,1)$       & $(1)-(1)$     &
                            $l_2 \neq -2,0,2$        \\
  \rule[-1.5ex]{0ex}{1.5ex}                                   &     &
                            $(0,l_2,3)$       & $(0)-(3)$     &
                            $l_2 \neq -1,0,3,4$      \\ \hline
  \rule[-1.5ex]{0ex}{4.5ex} $\mbox{\fraktw sl}(3\,|\,2)$      & $6$ &
                            $(0,0,l_3,0)$     & $(0,0)-(0)$   &
                            $l_3 \neq -2,-1,0,1$     \\ \hline
  \rule[-1.5ex]{0ex}{4.5ex} $\mbox{\fraktw osp}(2\,|\,4)$     & $4$ &
                            $(l_1,1,0)$       & $(1,0)$       &
                            $l_1 \neq 0,2,4,6$       \\ \hline
  \rule[-1.5ex]{0ex}{4.5ex} $\mbox{\fraktw osp}(2\,|\,6)$     & $6$ &
                            $(l_1,0,0,0)$     & $(0,0,0)$     &
                            $l_1 \neq 0,1,2,4,5,6$   \\ \hline
 \end{tabular}
 } \]
\end{table}

\vfill

\pagebreak

For type~II \LSA s, the analysis can be carried out along similar lines.
To begin with, we exclude the series $\, B(0\,|\,n) = \mbox{\fraktw osp}%
(1\,|\,2n)$ $(n \geq 1)$, since it does not provide any $64$-dimensional
irreducible representations. This can be derived from the remarkable fact
\cite{ris} that the irreducible representations of the type~II \LSA\
$\, B(0\,|\,n) = \mbox{\fraktw osp}(1\,|\,2n) \,$ (which by the way is
the only one for which all irreducible representations are typical) are
in one-to-one correspondence with those irreducible representations of the
ordinary simple Lie algebra $\, B_n = \mbox{\fraktw so}(2n+1) \,$ for which
the last Dynkin label, i.e., the coefficient $l_n$ associated with the short
simple root, is even -- a correspondence that can be represented graphically
in the form
\begin{equation}
 \begin{picture}(148,20)
  \thicklines
  \put(0,3){\makebox(35,0){{\large dim} {\Large (}}}
  \put(44,3){\circle{14}}
  \put(86,3){\circle{14}}
  \put(128,3){\circle*{14}}
  \put(44,18){\makebox(0.4,0.6){$l_1$}}
  \put(86,18){\makebox(0.4,0.6){$l_2$}}
  \put(128,18){\makebox(0.4,0.6){$l_n$}}
  \multiput(51,3)(5,0){6}{\line(1,0){3}}
  \put(92,0){\line(1,0){30}}
  \put(92,6){\line(1,0){30}}
  \put(113,3){\line(-1,1){10}}
  \put(113,3){\line(-1,-1){10}}
  \put(44,-7){$\underbrace{~~~~~~~~~~}_{n-1}$}
  \put(134,3){\makebox(10,0){{} {\Large )}}}
 \end{picture}~=~
 \begin{picture}(148,20)
  \thicklines
  \put(0,3){\makebox(35,0){{\large dim} {\Large (}}}
  \put(44,3){\circle{14}}
  \put(86,3){\circle{14}}
  \put(128,3){\circle{14}}
  \put(44,18){\makebox(0.4,0.6){$l_1$}}
  \put(86,18){\makebox(0.4,0.6){$l_2$}}
  \put(128,18){\makebox(0.4,0.6){$l_n$}}
  \multiput(51,3)(5,0){6}{\line(1,0){3}}
  \put(92,0){\line(1,0){30}}
  \put(92,6){\line(1,0){30}}
  \put(113,3){\line(-1,1){10}}
  \put(113,3){\line(-1,-1){10}}
  \put(44,-7){$\underbrace{~~~~~~~~~~}_{n-1}$}
  \put(134,3){\makebox(10,0){{} {\Large )}}}
 \end{picture}~.
\vspace{5mm}
\end{equation}
Note that there is no change in the Dynkin labels, so that according to
the integrality condition on the Dynkin label (\ref{eq:DYNL3}), $l_n$ must
be even, since for the $B(0,n)$ series, $s = n$, $\alpha_n^0 = 2\alpha_n \,$
and $\, l_n^{\,0} = {1 \over 2} \, l_n$. But it is known from evaluation of
the standard Weyl dimension formula that the only $64$-dimensional irreducible
representations of the $B_n$-series occur for $\, B_1 = \mbox{\fraktw so}(3)$,
with highest weight $63$, for $\, B_2 = \mbox{\fraktw so}(5)$, with highest
weight $(1,3)$, and for $\, B_6 = \mbox{\fraktw so}(13)$, with highest weight
$(0,0,0,0,0,1)$. For the remaining type~II \LSA s, we argue case by case, as
follows.
\begin{itemize}
 \item The series $\, B(m\,|\,n) = \mbox{\fraktw osp}(2m+1\,|\,2n) \,$
       with $\, m,n \geq 1 \,$: \\[1mm]
       For $\, \cg = \mbox{\fraktw osp}(2m+1\,|\,2n)$, we have
       $\, \cg_{\bar{0}} = \mbox{\fraktw so}(2m+1) \oplus
       \mbox{\fraktw sp}(2n)$, $r = m + n$, $s = n \,$ and
       $\, N_1 = (2m + 1)n$, so equation (\ref{eq:DIMFOR7})
       takes the form
       \begin{equation}
        \begin{array}{rcl}
         \dim V_\Lambda~=~2^{(2m+1)n} \!\!\!
          &\times&\!\!\! d_{\mbox{\ssfrak{sp}}(2n)}
                            (l_1,\ldots,l_{n-1},\tilde{l}_n^0) \\[1mm]
          &\times&\!\!\! d_{\mbox{\ssfrak{so}}(2m+1)}
                            (l_{n+1},\ldots,l_{n+m-1},l_{n+m})~,
        \end{array}
       \end{equation}
       where
       \begin{equation}
        l_n^{\,0}~=~l_n \, - \left( l_{n+1} \, + \ldots + \, l_{n+m-1} \, + \,
                            {\textstyle {1 \over 2}} \, l_{n+m} \right) \, ,
       \end{equation}
       and
       \begin{equation}
        \tilde{l}_n^{\,0}~=~l_n^{\,0} \, - \, m \, - \,
                            {\textstyle {1 \over 2}}~.
       \vspace{1mm}
       \end{equation}
       If $\, l_n^{\,0} < m$, write $\, l_n^{\,0} = k-1 \,$ where $\, 1 \leq
       k \leq m$; then the supplementary conditions \cite[pp.\ 251/252]{cor}
       require that
       $$
        l_{n+k}~=~\ldots~=~l_{n+m}~=~0~,
       $$
       and this forces $\, \Lambda + \rho \,$ to be orthogonal to the odd
       root $\, \epsilon_n^1 - \epsilon_k^2 \,$ \cite[pp.\ 513-521]{cor}.
       Similarly, if $\, l_n^{\,0} = m \,$ and we require in addition that
       $l_{n+m} = 0$, then $\, \Lambda + \rho \,$ will be orthogonal to the
       odd root $\, \epsilon_n^1 + \epsilon_m^2 \,$ \cite[pp.\ 513-521]{cor}.
       In both cases, this implies that the representation of $\cg$
       characterized by the highest weight $\Lambda$ is atypical.
       Thus we may assume that $\, l_n^{\,0} \geq m \,$ and use the
       monotonicity property (\ref{eq:MONOTO2}), distinguishing two
       cases:
 \begin{description}
  \item[$l_n^{\,0} > m \,$:] In this case,
        \begin{eqnarray*}
         \dim V_\Lambda \!\!
         &\geq&\!\! 2^{(2m+1)n} \;
                    d_{\mbox{\ssfrak{sp}}(2n)}(0,\ldots,0,
                                              {\textstyle {1 \over 2}}) \;
                    d_{\mbox{\ssfrak{so}}(2m+1)}(0,\ldots,0,0)          \\[1mm]
         & =  &\!\! 2^{2mn} \, {2n+1 \choose n}~.
        \end{eqnarray*}
  \item[$l_n^{\,0} = m \,$:] In this case,
        \begin{eqnarray*}
         \dim V_\Lambda \!\!
         &\geq&\!\! 2^{(2m+1)n} \;
                    d_{\mbox{\ssfrak{sp}}(2n)}(0,\ldots,0,
                                              - \, {\textstyle {1 \over 2}}) \;
                    d_{\mbox{\ssfrak{so}}(2m+1)}(0,\ldots,0,1)          \\[1mm]
         & =  &\!\! 2^{m(2n+1)}~.
        \end{eqnarray*}
 \end{description}
       In both cases, we conclude that $\, \dim V_\Lambda \,$ will exceed
       $64$ except when \linebreak $m = 1 \,$ and $\, n \leq 2 \,$ or when
       $\, m \leq 2 \,$ and $\, n = 1$.
 \item The series $\, D(m\,|\,n) = \mbox{\fraktw osp}(2m\,|\,2n) \,$
       with $\, m \geq 2 \,$ and $\, n \geq 1 \,$: \\[1mm]
       For $\, \cg = \mbox{\fraktw osp}(2m\,|\,2n)$, we have
       $\, \cg_{\bar{0}} = \mbox{\fraktw so}(2m) \oplus
       \mbox{\fraktw sp}(2n)$, $r = m + n$, $s = n \,$ and
       $\, N_1 = 2mn$, so equation (\ref{eq:DIMFOR7})
       takes the form
       \begin{equation}
        \begin{array}{rcl}
         \dim V_\Lambda~=~2^{2mn} \!\!\!
         &\times&\!\!\! d_{\mbox{\ssfrak{sp}}(2n)}
                          (l_1,\ldots,l_{n-1},\tilde{l}_n^0) \\[1mm]
         &\times&\!\!\! d_{\mbox{\ssfrak{so}}(2m)}
                          (l_{n+1},\ldots,l_{n+m-2},l_{n+m-1},l_{n+m})~,
        \end{array}
       \end{equation}
       where
       \begin{equation}
        l_n^{\,0}~=~l_n \, - \left( l_{n+1} \, + \ldots + \, l_{n+m-2} \, + \,
                             {\textstyle {1 \over 2}}
                             \left( l_{n+m-1} + l_{n+m} \right) \right) \, ,
       \end{equation}
       and
       \begin{equation}
        \tilde{l}_n^{\,0}~=~l_n^{\,0} \, - \, m~.
       \vspace{1mm}
       \end{equation}
       If $\, l_n^{\,0} < m$, write $\, l_n^{\,0} = k-1 \,$ where $\, 1 \leq
       k \leq m$; then the supplementary conditions \cite[pp.\ 251/252]{cor}
       require that
       \begin{eqnarray*}
        &l_{n+k}~=~\ldots~=~l_{n+m}~=~0 \qquad
         \mbox{if $\, l_n^{\,0} < m-1$}~,& \\
        &l_{n+m-1}~=~l_{n+m} \qquad
         \mbox{if $\, l_n^{\,0} = m-1$}~,&
       \end{eqnarray*}
       and this forces $\, \Lambda + \rho \,$ to be orthogonal to the odd
       root $\, \epsilon_n^1 - \epsilon_k^2 \,$ \cite[pp.\ 525-532]{cor}.
       Similarly, if $\, l_n^{\,0} = m \,$ and we require in addition that
       $\, l_{n+m-1} = 0 \,$ and $\, l_{n+m} = 0$, then $\, \Lambda + \rho \,$
       will be orthogonal to the odd root $\, \epsilon_n^1 + \epsilon_{m-1}^2$
       \cite[pp.\ 525-532]{cor}. In both cases, this implies that the
       representation of $\cg$ characterized by the highest weight
       $\Lambda$ is atypical. Thus we may assume that $\, l_n^{\,0}
       \geq m \,$ and use the monotonicity property (\ref{eq:MONOTO2}),
       distinguishing two cases:
 \begin{description}
  \item[$l_n^{\,0} > m \,$:] In this case,
        \begin{eqnarray*}
         \dim V_\Lambda \!\!
         &\geq&\!\! 2^{2mn} \;
                    d_{\mbox{\ssfrak{sp}}(2n)}(0,\ldots,0,1) \;
                    d_{\mbox{\ssfrak{so}}(2m)}(0,\ldots,0,0,0)          \\[1mm]
         & =  &\!\! 2^{2mn+1} \, {1 \over n} \, {2n+1 \choose n-1}~.
        \end{eqnarray*}
  \item[$l_n^{\,0} = m \,$:] In this case, $l_{n+m-1} > 0 \,$ and
        \begin{eqnarray*}
         \dim V_\Lambda \!\!
         &\geq&\!\! 2^{2mn} \;
                    d_{\mbox{\ssfrak{sp}}(2n)}(0,\ldots,0,0) \;
                    d_{\mbox{\ssfrak{so}}(2m)}(0,\ldots,0,1,0)          \\[1mm]
         & =  &\!\! 2^{m(2n+1)-1}~,
        \end{eqnarray*}
        \hspace*{6mm} or $\, l_{n+m} > 0 \,$ and
        \begin{eqnarray*}
         \dim V_\Lambda \!\!
         &\geq&\!\! 2^{2mn} \;
                    d_{\mbox{\ssfrak{sp}}(2n)}(0,\ldots,0,0) \;
                    d_{\mbox{\ssfrak{so}}(2m)}(0,\ldots,0,0,1)          \\[1mm]
         & =  &\!\! 2^{m(2n+1)-1}~.
        \end{eqnarray*}
 \end{description}
       In both cases, we conclude that $\, \dim V_\Lambda \,$ will exceed
       $64$ except when \linebreak $m = 2 \,$ and $\, n = 1$.
 \item The family $\, D(2\,|\,1\,;\alpha)
       = \mbox{\fraktw osp}(4\,|\,2\,;\alpha) \,$
       with $\, \alpha \neq 0,-1,\infty \,$: \\[1mm]
       For $\, \cg = \mbox{\fraktw osp}(4\,|\,2\,;\alpha)$, we have
       $\, \cg_{\bar{0}} = \mbox{\fraktw su}(2) \oplus \mbox{\fraktw su}(2)
       \oplus \mbox{\fraktw su}(2)$, $r = 3$, $s = 1 \,$ and $\, N_1 = 4$,
       so equation (\ref{eq:DIMFOR7}) takes the form
       \begin{eqnarray}
        \dim V_\Lambda \!\!
        &=&\!\! 16 \; d_{\mbox{\ssfrak{su}}(2)}(\tilde{l}_1^{\,0}) \;
                      d_{\mbox{\ssfrak{su}}(2)}(l_2) \;
                      d_{\mbox{\ssfrak{su}}(2)}(l_3)                    \\[1mm]
        &=&\!\! 16 \; (1 + \tilde{l}_1^{\,0}) \, (1 + l_2) \, (1 + l_3)~,
       \end{eqnarray}
       where
       \begin{equation}
        l_1^{\,0}~=~l_1 \, - \, {\textstyle {1 \over 2}}
                                \left( l_2 + l_3 \right) \, ,
       \end{equation}
       and
       \begin{equation}
        \tilde{l}_1^{\,0}~=~l_1^{\,0} \, - \, 2~.
       \vspace{1mm}
       \end{equation}
       If $\, l_1^{\,0} < 2$, the supplementary conditions
       \cite[pp.\ 251/252]{cor} require that
       \begin{eqnarray*}
        &l_2~=~l_3~=~0 \qquad
         \mbox{if $\, l_1^{\,0} = 0$}~,& \\
        &\alpha \left( l_3 + 1 \right)~=~l_2 + 1 \qquad
         \mbox{if $\, l_1^{\,0} = 1$}~,&
       \end{eqnarray*}
       and this forces $\, \Lambda + \rho \,$ to be orthogonal to the
       simple odd root $\, \alpha_1 \,$ in the first case and to the odd root
       $\, \alpha_1 + \alpha_2 \,$ in the second case \cite[pp.\ 532-537]{cor},
       which implies that the representation of $\cg$ characterized by the
       highest weight $\Lambda$ is atypical. Thus we may assume that
       $\, l_1^{\,0} \geq 2.$
 \item The algebra $\, F(4) \,$: \\[1mm]
       For $\, \cg = F(4)$, we have $\, \cg_{\bar{0}} = \mbox{\fraktw su}(2)
       \oplus \mbox{\fraktw so}(7)$, $r = 4$, $s = 1 \,$ and $\, N_1 = 8$,
       so equation (\ref{eq:DIMFOR7}) takes the form
       \begin{eqnarray}
        \dim V_\Lambda \!\!
        &=&\!\! 256 \; d_{\mbox{\ssfrak{su}}(2)}(\tilde{l}_1^{\,0}) \;
                       d_{\mbox{\ssfrak{so}}(7)}(l_4,l_3,l_2)           \\[1mm]
        &=&\!\! 256 \; (1 + \tilde{l}_1^{\,0}) \;
                       d_{\mbox{\ssfrak{so}}(7)}(l_4,l_3,l_2)~,
       \end{eqnarray}
       where
       \begin{equation}
        l_1^{\,0}~=~{\textstyle {1 \over 3}}
                    \left( 2 l_1 - 3 l_2 - 4 l_3 - 2 l_4 \right) \, ,
       \end{equation}
       and
       \begin{equation}
        \tilde{l}_1^{\,0}~=~l_1^{\,0} \, - \, 4~.
       \vspace{1mm}
       \end{equation}
       If $\, l_1^{\,0} < 4$, the supplementary conditions
       \cite[pp.\ 251/252]{cor} require that $\, l_1^{\,0} \neq 1 \,$ and
       \begin{eqnarray*}
        &l_2~=~l_3~=~l_4~=~0 \qquad
         \mbox{if $\, l_1^{\,0} = 0$}~,& \\
        &l_2~=~l_4~=~0 \qquad
         \mbox{if $\, l_1^{\,0} = 2$}~,& \\
        &l_2~=~2 l_4 + 1 \qquad
         \mbox{if $\, l_1^{\,0} = 3$}~,&
       \end{eqnarray*}
       and this forces $\, \Lambda + \rho \,$ to be orthogonal to the
       simple odd root $\, \alpha_1 \,$ in the first case, to the odd root
       $\, \alpha_1 + \alpha_2 + \alpha_3 \,$ in the second case and to the
       odd root $\, \alpha_1 + \alpha_2 + \alpha_3 + \alpha_4 \,$ in the third
       case \cite[pp.\ 537-541]{cor}, which implies that the representation of
       $\cg$ characterized by the highest weight $\Lambda$ is atypical. Thus
       we may assume that $\, l_1^{\,0} \geq 4 \,$ and deduce that the
       the dimension of any typical representation of $F(4)$ is a
       multiple of $256$.
 \item The algebra $\, G(3) \,$: \\[1mm]
       For $\, \cg = G(3)$, we have $\, \cg_{\bar{0}} = \mbox{\fraktw su}(2)
       \oplus G_2$, $r = 3$, $s = 1 \,$ and $\, N_1 = 7$, so equation
       (\ref{eq:DIMFOR7}) takes the form
       \begin{eqnarray}
        \dim V_\Lambda \!\!
        &=&\!\! 128 \; d_{\mbox{\ssfrak{su}}(2)}(\tilde{l}_1^{\,0}) \;
                       d_{G_2}(l_3,l_2)                                 \\[1mm]
        &=&\!\! 128 \; (1 + \tilde{l}_1^{\,0}) \; d_{G_2}(l_3,l_2)~,
       \end{eqnarray}
       where
       \begin{equation}
        l_1^{\,0}~=~{\textstyle {1 \over 2}}
                    \left( l_1 - 2 l_2 - 3 l_3 \right) \, ,
       \end{equation}
       and
       \begin{equation}
        \tilde{l}_1^{\,0}~=~l_1^{\,0} \, - \, {\textstyle {7 \over 2}}~.
       \vspace{1mm}
       \end{equation}
       If $\, l_1^{\,0} < 3$, the supplementary conditions
       \cite[pp.\ 251/252]{cor} require that $\, l_1^{\,0} \neq 1 \,$ and
       \begin{eqnarray*}
        &l_2~=~l_3~=~0 \qquad \mbox{if $\, l_1^{\,0} = 0$}~,& \\
        &l_2~=~0 \qquad \mbox{if $\, l_1^{\,0} = 3$}~,&
       \end{eqnarray*}
       and this forces $\, \Lambda + \rho \,$ to be orthogonal to the
       simple odd root $\, \alpha_1 \,$ in the first case and to the
       odd root $\, \alpha_1 + \alpha_2 + \alpha_3 \,$ in the second
       case \cite[pp.\ 542-545]{cor}. Similarly, if $\, l_1^{\,0} = 3 \,$
       and we require in addition that $\, l_2 = 0 \,$ and $\, l_3 = 0$,
       then $\, \Lambda + \rho \,$ will be orthogonal to the odd root
       $\, \alpha_1 + 3 \alpha_2 + \alpha_3 \,$ \cite[pp.\ 542-545]{cor}.
       In both cases, this implies that the representation of $\cg$
       characterized by the highest weight $\Lambda$ is atypical.
       Thus we may assume that $\, l_1^{\,0} \geq 3 \,$ and deduce
       that the dimension of any typical representation of $G(3)$ is
       a multiple of $64$; moreover, the only candidate of dimension
       equal to $64$ ($l_1 = 6$, \mbox{$l_2 = 0$,} $l_3 = 0$) is
       excluded, because it is atypical.
\end{itemize}
With these restrictions, it is now an easy exercise to write down the
highest weights of all irreducible representations of $\cg_{\bar{0}}$
of the correct dimension and to eliminate all candidates that fail to
satisfy the typicality conditions; the result is presented in Table 4.
Note that in the family $\, D(2\,|\,1;\alpha) = \mbox{\fraktw osp}%
(4\,|\,2\,;\alpha)$, the parameter $\alpha$ remains unspecified: it can
take any complex value except $0$, $-1$, $\infty$ and a few other rational
numbers that must be excluded in order to guarantee that the representation
is indeed typical, and its choice has no influence on the dimension of the
representation.

\vfill

\begin{table}[ht] \label{tb:CRTYPE2}
 \caption{Typical codon representations of type~II \LSA s}
 \vspace{3ex}
 \[ \mbox{
 \begin{tabular}{|c|c|c|c|} \hline
  \rule{0ex}{3ex}           Lie            &
                            Highest Weight & Highest Weight     & Typicality \\
  \rule[-1.5ex]{0ex}{1.5ex} Superalgebra   &
                            of $\cg$       & of $\cg_{\bar{0}}$ & Condition
  \\ \hline\hline
  \rule[-1.5ex]{0ex}{4.5ex} $\mbox{\fraktw osp}(3\,|\,2)$        &
                            $({17 \over 2},15)$ & $(1)-(15)$     & \\ \hline
  \rule[-1.5ex]{0ex}{4.5ex} $\mbox{\fraktw osp}(5\,|\,2)$        &
                            $({5 \over 2},0,1)$ & $(2)-(0,1)$    & \\ \hline
  \rule[-1.5ex]{0ex}{4.5ex} $\mbox{\fraktw osp}(3\,|\,4)$        &
                            $(0,{5 \over 2},3)$ & $(0,1)-(3)$    & \\ \hline
  \rule[-1ex]{0ex}{4ex}   & $({1 \over 2} \, (5\alpha + 5),0,0)$ &
                            $(5)-(0)-(0)$ &
                            $\alpha \neq - {5 \over 3}, - {3 \over 5}$ \\
  \rule[-1ex]{0ex}{4ex}   & $({1 \over 2} \, (3\alpha + 4),1,0)$ &
                            $(3)-(1)-(0)$ &
                            $\alpha \neq - 4, - {4 \over 3}$           \\
  \rule[-1ex]{0ex}{4ex}   & $({1 \over 2} \, (4\alpha + 3),0,1)$ &
                            $(3)-(0)-(1)$ &
                            $\alpha \neq - {1 \over 4}, - {3 \over 4}$ \\
  \rule[-1ex]{0ex}{4ex}
   \raisebox{1.5ex}[-1.5ex]{$\mbox{\fraktw osp}(4\,|\,2\,;\alpha)$}
                          & $({1 \over 2} \, (3\alpha + 3),1,1)$ &
                            $(2)-(1)-(1)$ &
                            $\alpha \neq 3, - {1 \over 3}$             \\
  \rule[-1ex]{0ex}{4ex}   & $({1 \over 2} \, (2\alpha + 5),3,0)$ &
                            $(2)-(3)-(0)$ &
                            $\alpha \neq - {5 \over 2}, {3 \over 2}$   \\
  \rule[-1.5ex]{0ex}{4.5ex}&$({1 \over 2} \, (5\alpha + 2),0,3)$ &
                            $(2)-(0)-(3)$ &
                            $\alpha \neq - {2 \over 5}, {2 \over 3}$   \\ \hline
 \end{tabular}
 } \]
\end{table}

\vfill \mbox{}

\pagebreak

\section{Conclusions and Outlook}

The main result of the present paper, the first in a sequence of two, is the
complete list of all typical codon representations (typical $64$-dimensional
irreducible representations) of basic classical \LSA s, presented in Table 3
and Table 4: we find $12$ basic classical \LSA s with a total of $18$ codon
representations that are essentially different (conjugate representations
are not regarded as essentially different). The analysis is based on the
classification of basic classical \LSA s and on their representation theory,
which are briefly reviewed in Sect.\ 2 and Sect.\ 3, respectively, in
particular on the Weyl-Kac dimension formula for typical representations.
As in the case of the ordinary Weyl dimension formula for irreducible
representations of ordinary simple Lie algebras, the dimension of the
representation grows with its highest weight, so that no algebra belonging
to any of the classical series will, from a certain rank upwards, admit
codon representations (or, more generally, non-trivial representations
of dimension $\leq 64$) at all. The main difficulty to be overcome was
to extend this monotonicity argument from ordinary simple Lie algebras
to basic classical \LSA s and to derive sufficiently sharp lower bounds
on dimensions of typical representations, in order to exclude the appearance
of algebras of higher rank. In this respect, the final results are more
stringent for basic classical \LSA s than they are for ordinary simple
Lie algebras.

On the other hand, it must be stressed that for atypical representations,
a general dimension formula is still not known, and this is a major obstacle
to performing a similar analysis for this kind of representations -- despite
the fact that there is no reason to regard typical representations as being
more important than atypical ones; see \cite[p.\ 258/259]{cor} for comments
on this matter. Similarly, the existence and classification of codon
representations of the strange classical \LSA s is an open problem.
In this sense, the analysis presented in the present paper is not
complete.

Despite these limitations, our investigation does provide a framework for
the subsequent investigation of branching schemes, the main goal being to
identify the ones that reproduce the standard genetic code. This analysis
will be performed in the forthcoming second paper of this series.
%The numerical doubling of typical codon representations of basic classical
%\LSA s as compared to codon representations of simple Lie algebras enlarges
%the chances of success for the following steps in this program.
%\begin{itemize}
% \item[1.)] We will indicate the branching schemes most similar to the
%            structure of the standard genetic code.
% \item[2.)] Hamiltonian operators (codon operators) will be constructed from
%            polynomial functions in the Lie generators of the superalgebra on
%            which the respective model is based. For the first evolutionary
%            steps it should bear a dynamical symmetry in just incorporating
%            Casimir operators of the corresponding subalgebras violating this
%            condition in the last step only.
% \item[3.)] Given the Hamiltonian which exhibits the correct spectral
%            degeneracy it will be investigated its physico-chemical relevance
%            as it may account for electrical polarities of the amino acids or
%            similar properties.
% \item[4.)] Another important question is whether a symmetry branching scheme
%            provides an explanation for some characteristics of the genetic
%            code like wobbling rules \cite{cri,juk1,juk2}, i.e., the
%            preference of family boxes and to the occurence of modifications
%            of the genetic code in mitochondrial DNA.
%\end{itemize}

\subsection*{Acknowledgments}

The authors would like to thank Prof.\ J.E.M.\ Hornos for his incentive and
support of the present project, Prof.\ A.~Sciarrino and Prof.\ P.~Jarvis for
clarifying correspondence on the representation theory of \LSA s and Prof.\
\mbox{A.\ Grishkov} for fruitful discussions.

\pagebreak

\end{document}